\documentclass[11pt, letterpaper]{article}
\usepackage[utf8]{inputenc}
\usepackage[margin=1.5cm]{geometry}
\usepackage{titlesec}
\usepackage{tabu}
\usepackage{longtable}
\usepackage{enumitem}
\usepackage{amssymb}
\usepackage{xcolor}
\newlist{selectlist}{itemize}{2}
\setlist[selectlist]{label=$\square$,leftmargin=*,noitemsep,topsep=0pt}

\usepackage{lmodern}

\usepackage{hyperref}
\hypersetup{
 colorlinks=true,
 linkcolor=blue,
 filecolor=magenta, 
 urlcolor=blue,
}

\usepackage{graphicx}

\titleformat{\section}[block]{\hspace{1em}\bfseries}{\thesection.}{0.5em}{} 
\titleformat{\subsection}[block]{\hspace{1em}}{\thesubsection}{0.5em}{}

\begin{document}
% Create the title block
\begin{flushleft}

\textbf{Article information}\\
\vskip 0.5cm
\textbf{\emph{Cultural gems linked open data: Mapping culture and intangible heritage in European cities}} \vskip 0.5cm
\textbf{Authors}\\ 
Sergio Consoli$^{a,*}$, Valentina Alberti$^a$, Cinzia Cocco$^a$, Francesco Panella$^a$, Valentina Montalto$^a$
\vskip 0.5cm
%Insert Affiliations
\textbf{Affiliations}\\ 
$^a$ European Commission, Joint Research Centre (JRC), Via E. Fermi 2749, I-21027 Ispra (VA), Italy. [\textit{name.surname}]@ec.europa.eu 
\vskip 0.5cm
%Insert Contact Email
%Include institutional email address of the corresponding author
\textbf{Corresponding author’s email address and Twitter handle}\\ 
Sergio Consoli \\
Email: sergio.consoli@ec.europa.eu\\
Twitter: @jattokatarratto
\vskip 0.5cm
%Insert Keywords
\textbf{Keywords}\\ 
ICT and Society; Social applications of the Semantic Web; Linked Open Data for Cultural Heritage; Cultural domain ontologies; Semantic Web content creation, annotation, and extraction; Ontology mapping, merging, and alignment; RDF data processing.
\vskip 0.5cm
%Insert Abstract
\textbf{Abstract}\\ 
The recovery and resilience of the cultural and creative sectors after the COVID-19 pandemic
is a current topic with priority for the European Commission. % assesses the current Work Plan for Culture
%is New topics that deserve attention
%should be prioritised
\emph{Cultural gems} is a crowdsourced web platform managed by the Joint Research Centre of the European Commission aimed at creating community-led maps as well as a common repository for cultural and creative places across European cities and towns. More than 130,000 physical locations and online cultural activities in more than 300 European cities and towns are currently tracked %mapped 
by the application. The main objective of \emph{Cultural gems} consists in raising a holistic vision of European culture, reinforcing a sense of belonging to a common European cultural space. 
%Rephrased for DIB
%\emph{Cultural gems} maps more than 130,000 physical places in over 300 European cities and towns, and since 2020 it also lists online cultural initiatives.
This data article describes the ontology developed for \emph{Cultural gems}, adopted to represent the domain of knowledge of the application %which is used to map cultural heritage in European cities 
by means of \emph{FAIR} (\emph{F}indable, \emph{A}ccessible, \emph{I}nteroperable, \emph{R}eusable) principles and following the paradigms of Linked Open Data (LOD). 
%To this end, we exploit a dataset 
We provide an overview of this dataset, % and methodology, % under current development, 
and describe the ontology model, along with the services %and tools 
used to access and consume the data.
\vskip 0.5cm
\textbf{Specifications table}\\
\vskip 0.3cm 
%
%\vskip6pt\noindent
%{\small\textbf{\textit{Please delete this line and everything above it before submitting your
%article, in addition to anything in [square brackets] below, including
%in the Specifications Table}}\vskip6pt\hrule\vskip12pt}
%
%{\fontsize{7.5pt}{9pt}\selectfont
%%%%
%\noindent\textbf{Specifications Table} 
%
%Every section of this table is mandatory. 
%Please enter information in the right-hand column and remove all the instructions
\begin{longtable}{|p{33mm}|p{124mm}|}
%\hline
%\endhead
\hline
%\endfoot
\textbf{Subject} & Arts and Humanities\\
\hline 
\textbf{Specific subject area} & Ontology engineering, knowledge management, and linked open data technologies to model the European cultural heritage \\
\hline
\textbf{Type of data} & %Graph \\ 
Text files \newline
Graphs \newline
OWL files\newline
RDF/XML files\newline
Turtle files \newline
Figures\\
\hline
\textbf{How the data were acquired} & Data was acquired by fetching \emph{Cultural gems} (CG), a public application \cite{ourinpress2023, CGapp} designed and managed by the EC Joint Research Centre (DG JRC) \cite{JRCpage}. It is a free and open-source web platform, crowdsourced, which maps cultural and creative venues in European cities.
\emph{Cultural gems} includes data on selected cultural venues from OpenStreetMap \cite{OpenStreetMap}, and information provided by European cities, universities, public and private organisations, and other individuals that can share their data, to allow them to visualise the information on enriched city maps. 
Data resources have been also linked using \textbf{owl:sameAs} axioms to other public ontologies: DBpedia \cite{DBpedia} and GeoNames \cite{GeoNames}. 
Data is constantly expanding \cite{JRCdataCG} and further alignments to other popular semantic datasets, freely accessible online and specialised in the cultural heritage domain, will follow. \newline
To design the linked open data model for \emph{Cultural gems} we have directly mapped the main definitions from classes defined in the CG platform. These are primarily structured according to the OpenStreetMap classification \cite{OpenStreetMap} and the ``concentric circles model of cultural industries'' \cite{Throsby2008}. We also adopt ontology design patterns (ODPs) principles \cite{gangemi2009ontology} %Gangemi2005262 
to design the ontology. As such, %that is, whenever possible 
we reuse ODPs already defined in other public ontologies as much as possible. 
We rely on the OPLa ontology \cite{Asprino2021} %Shimizu201823 
to identify these ODPs for an intuitive understanding of the overall class mapping of the ontology.\\
\hline 
\textbf{Data format} & 
Raw\newline
Filtered
\\ 
\hline
\textbf{Description of 
data collection} & Data were collected from the \emph{Cultural gems} application and 
transformed to individuals of the produced ontology through an ETL (Extract-Transform-Load) nightly job. The code has been developed using the Python programming language.
The resulting data ontology currently accounts for around 2.9M triples \cite{JRCdataCG}. It is available in both RDF/XML and Turtle file formats. The reference namespace for the data is: 
\href{https://jeodpp.jrc.ec.europa.eu/ftp/jrc-opendata/CC-COIN/cultural-gems/ontology-individuals/cultural-gems-resources.owl}{https://culturalgems.jrc.ec.europa.eu/resource/}. 
Instead, the reference namespace for the ontology definition is: 
\href{https://jeodpp.jrc.ec.europa.eu/ftp/jrc-opendata/CC-COIN/cultural-gems/ontology-definition/cultural-gems.owl}{https://culturalgems.jrc.ec.europa.eu/ontology/cultural-gems/}, for a total of 67 classes. 
%accounting so far to an overall of 67 classes. 
\newline
%Commonly-used style guidelines for labeling and representing ontology definitions and resources have been employed in our ontology engineering task. 
In our ontology engineering work, we have adopted commonly-accepted style rules for identifying and describing ontology objects. %Specifically, ontology data names within the ontology data have been represented in lowercase, substituting eventual space characters with dashes. Ontology definition class names, instead, have been represented using uppercase. 
Specifically, data names inside the ontology have been expressed in lowercase, with dashes used in place of any potential space characters. Class names from ontology definitions, likewise, have been written in uppercase. \\
\hline 
\textbf{Data source location} & 
\noindent
%$\bullet$ 
European Commission, Joint Research Centre (JRC)\newline
%$\bullet$ 
Via E. Fermi 2749, I - 21027 \newline
%$\bullet$ 
Ispra, Varese\newline
%$\bullet$ 
Italy\\
\hline 
\hypertarget{target1}
{\textbf{Data accessibility}} & 
Source ontology definition and data files are available in RDF/XML and Turtle formats \cite{JRCdataCG} within the Joint Research Centre Data Catalogue \cite{JRCDataCatalogue} at the following permanent location:
\url{https://data.jrc.ec.europa.eu/dataset/9ee32efe-af81-48e4-8ad6-a0db06802e03}. These ontologies have been also released within the European Data portal \cite{EUDataPortal}, the official data repository for European data, at the following permanent location: \url{https://data.europa.eu/data/datasets/9ee32efe-af81-48e4-8ad6-a0db06802e03}.\newline
We enable %access %allow user-friendly access to allow 
the interested community %citizens, developers, and companies 
to access and consume the produced data and ontology under Creative Commons Attribution 4.0 International (CC BY 4.0) license. % \cite{CCBY4}.
\newline
\vskip 0.2cm
Repository name: Joint Research Centre Data Catalogue\newline
Data identification number: \url{http://data.europa.eu/89h/9ee32efe-af81-48e4-8ad6-a0db06802e03}\newline
Direct URL to data: \url{https://data.jrc.ec.europa.eu/dataset/9ee32efe-af81-48e4-8ad6-a0db06802e03}\newline
\vskip 0.2cm
Repository name: European Data portal\newline
Data identification number: \url{http://data.europa.eu/88u/dataset/9ee32efe-af81-48e4-8ad6-a0db06802e03}\newline
Direct URL to data: \url{https://data.europa.eu/data/datasets/9ee32efe-af81-48e4-8ad6-a0db06802e03}\newline
\vskip 0.2cm

Data are freely available to the public and can be downloaded without any login details by selecting the appropriate URL in one of the repositories above. In particular, for the sake of clarity, we specify below the URL of each of the released data source:

\begin{itemize}
 \item Cultural gems ontology definition (RDF/OWL): \url{https://jeodpp.jrc.ec.europa.eu/ftp/jrc-opendata/CC-COIN/cultural-gems/ontology-definition/cultural-gems.owl}
 \item Cultural gems ontology definition - no imports (RDF/OWL): \url{https://jeodpp.jrc.ec.europa.eu/ftp/jrc-opendata/CC-COIN/cultural-gems/ontology-definition/cultural-gems-skeleton.owl}
 \item Cultural gems ontology individuals (RDF/OWL): \url{https://jeodpp.jrc.ec.europa.eu/ftp/jrc-opendata/CC-COIN/cultural-gems/ontology-individuals/cultural-gems-resources.owl}
 \item Cultural gems ontology individuals (Turtle): \url{https://jeodpp.jrc.ec.europa.eu/ftp/jrc-opendata/CC-COIN/cultural-gems/ontology-individuals/cultural-gems-resources.ttl}
\end{itemize}

We also enable direct access to the data and ontology by means of SPARQL queries
on CELLAR \cite{Cellar}, the Publications Office's common repository of metadata and content, by using its REST APIs SPARQL endpoint services. %\footnote{CELLAR's SPARQL endpoint service: \url{https://data.europa.eu/data/datasets/sparql-cellar-of-the-publications-office}}.
The REST web service to access the dedicated CELLAR SPARQL endpoint is \url{https://publications.europa.eu/webapi/rdf/sparql}.\\ 
\hline 
%\textbf{Related research\newline article} & 
%V. Alberti, C. Cocco, S. Consoli, V. Montalto, F. Panella, Ontology engineering to model the European cultural heritage: The case of Cultural gems, Lecture Notes in Networks and Systems. In Press.\\
%\hline 
\end{longtable}

%%% 
\vskip 0.5cm

\textbf{Value of the data}\\

\begin{itemize}
\itemsep=0pt
\parsep=0pt
\item[$\bullet$] %This data creates a unique crowdsourced repository of cultural and creative places from people and organisations across Europe.
Using information from individuals and organizations all around Europe, this data creates a unique crowdsourced archive of cultural and creative locations.

\item[$\bullet$] %The released ontology represents a digital resource %, accompanied by several services, 
%to help local authorities and individuals who work or are interested in the domain of arts and culture,
%%, or are just interested, in the artistic and cultural domains, 
%to promote these sectors within their municipalities.
%
The released ontology serves as a digital promotion tool for local government entities and anyone working within the arts and culture sector.

\item[$\bullet$] The dataset may be beneficial for people concerned with research and development tools in the cultural heritage field.

\item[$\bullet$] This data helps to interconnect practitioners and researchers exploiting the cultural heritage sector. % contributing to the EU-wide project of culture and creativity promotion.

\item[$\bullet$] Local authorities, research organizations, municipalities, and European institutions %private and public institutions in Europe, 
may be beneficial in the use of this data for the development of cultural information services and systems.

\item[$\bullet$] An additional value is that the ontology is sharable, extensible, and easily reusable. Resources can be described jointly with other linked cultural datasets contributed by other communities, enabling interoperability at both syntactic and semantic levels. % with further popular ontologies specialised in the cultural heritage field.

\end{itemize}
\vskip 0.5cm

\textbf{Objective}\\
\vskip0.3cm 

The main objective of the released ontology consists of improving the data infrastructure of the European Commission's \emph{Cultural gems} application. The objective is to allow proper management of cultural properties, events, stories, geo-locations (points, lines, polygons, ...), intangible heritage, and the time dimension (one-time, recurrent). The ontology model further enhances information accessibility and interoperability.

The goal is to support and encourage European organisations and individuals to interact with the application to build a unique database of cultural and creative locations. %create a unique repository of cultural and creative places, providing a
%common vision of European culture to strengthen a sense of identity within a single European cultural realm. 
We want to support a shared awareness of European culture to reinforce the perception of one European cultural domain. 
For this purpose, the released ontology \cite{JRCdataCG} exploits the large capabilities of Semantic Web technologies %ous potential of Semantic technologies and Linked Open Data (LOD) %\cite{Hyvnen2020187}, %Heath20111 
to connect with the various datasets and ontologies existing in the cultural domain \cite{Hyvnen2020187}.
%to facilitate the interconnection with disparate datasets and ontologies in the cultural heritage field \cite{Hyvnen2020187}, %Stasinopoulou2007165,Popovici2011105 such as Europeana\footnote{Euopeana: 
%\url{https://www.europeana.eu/en}} \cite{europeana}, %and the Knowledge Graph of the Italian Cultural Heritage (ArCo)\footnote{%ArCo - Knowledge Graph of the Italian Cultural Heritage: 
%\url{http://dati.beniculturali.it/arco/}} %http://wit.istc.cnr.it/arco/index.php?lang=en}} 
%\cite{Carriero201936,arcoJournal}, 
%using metadata standards, and increasing semantic interoperability of the application.
We aim at enabling the adoption of several additional services and tools that we are currently building on top of CG to increase the final user experience. %the interoperability and usability of the application.

\vskip 0.5cm

\textbf{Data description}\\
\vskip0.3cm 

The primary classes employed in the platform, whose present taxonomy is roughly based on the ``concentric circles model of cultural industries'' \cite{Throsby2008}, have been utilized for developing the released \emph{Cultural gems} ontology.
The ontology seeks to properly represent the cultural data of the application. 
%The ontology aims at closely modelling %wide 
%this cultural heritage data. % in the application. 
%The categories are mainly organised to match also the OpenStreetMap categorization \cite{OpenStreetMap} %\footnote{OpenStreetMap features description: \url{https://wiki.openstreetmap.org/wiki/Map_features}} 
%relevant %and adopted in
%to our mapping purposes for both interoperability and clarity purposes. 
% related to European cities. 
For interoperability and simplicity, the classes have been mostly structured to reflect the OpenStreetMap categorization \cite{OpenStreetMap}, useful to our mapping requirements.

%The goal consists of building an ontology that is compatible, and aligned whenever possible, with existing ontologies in the related domain, that are used as a de facto standard for representing cultural heritage data. These include, for example, the already mentioned Europeana and ArCo ontologies, and services like the Hellenic Aggregator of Digital Cultural Content\footnote{%SearchCulture.gr:
%\url{https://www.searchculture.gr/aggregator/portal/}}, among others, for linking and aggregating the generated application data with the various cultural heritage LOD available online. %in the Semantic Web.

\vskip0.3cm

%Furthermore, 
Our designed ontology model reuses several properties and classes of ArCo \cite{arcoJournal}, the knowledge graph of the Italian cultural heritage, 
%the ArCo network of ontologies %\footnote{ArCo - Knowledge Graph of the Italian Cultural Heritage: \url{http://dati.beniculturali.it/arco/}} %http://wit.istc.cnr.it/arco/index.php?lang=en}} \cite{Carriero201936,arcoJournal} 
utilizing \emph{owl:imports} relationships.
%\textbf{d2rq:PropertyBridge}
ArCo is used as the top-level taxonomy of the \emph{Cultural gems} ontology. CG concepts are modelled as subclasses of ArCo in RDF/OWL standard format. 
% The main classes of our \emph{Cultural gems} ontology are mapped to RDF/OWL as subclasses of the top-level hierarchy of ArCo.
% ontology, 
In particular, we reuse the ArCo \emph{:TangibleCulturalProperty} and \emph{:IntangibleCulturalProperty} classes, defined as subclasses of \emph{:CulturalProperty}. Furthermore, the \emph{:TangibleCulturalProperty} concept is refined 
into the \emph{:MovableCulturalProperty} and \emph{:ImmovableCulturalProperty} classes. In addition, we further adopt the 
\emph{:HistoricOrArtisticProperty}, \emph{:ArchaeologicalProperty}, and \emph{:MusicHeritage} cultural classes from ArCo.
%Other specific types of cultural properties we reuse are: \emph{:ArchaeologicalProperty}, \emph{:HistoricOrArtisticProperty}, and \emph{:MusicHeritage}. 
%The cultural events module, which extends the Cultural-ON ontology \cite{Lodi2017}, has been also used to map cultural events and exhibitions involving a cultural property. 
Exhibitions and events involving a cultural property have also been represented using the cultural events ontology of ArCo, which expands Cultural-ON \cite{Lodi2017}.
\begin{figure}[t!]
\begin{center}
 \includegraphics[width=0.99\linewidth]{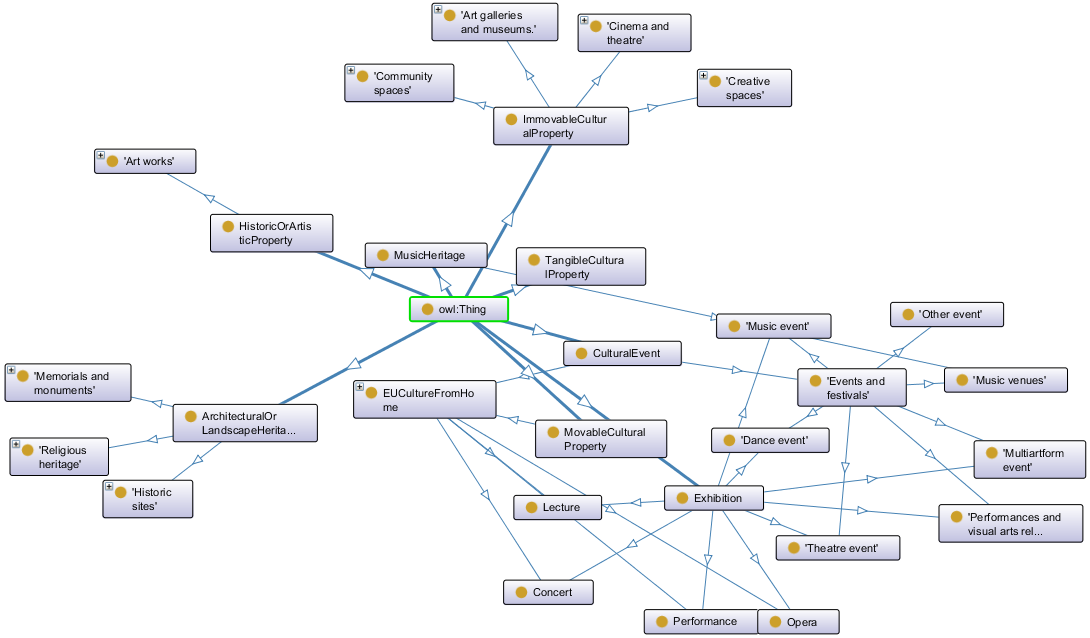}
\caption{Main classes of the \emph{Cultural gems} ontology.}
	\label{onto-mapping}
\end{center}
\end{figure}

\vskip0.3cm

Figure \ref{onto-mapping} illustrates how the top classes of \emph{Cultural gems} are connected to these ones. %reused classes definitions,
%The main classes of \emph{Cultural gems} are mapped to those class definitions, as illustrated in Figure \ref{onto-mapping}. 
The primary class hierarchy of CG is described as follows. %in the consists of the following classes: 
%\begin{itemize}
 %\item 

 \vskip0.3cm
 
 The \emph{EUCultureFromHome} class  % - Class representing cultural initiatives in European cities accessible online. For example, travel restrictions and social distancing might be limiting the possibility to visit venues and to taste cultural fragrances of European cities and towns in person. Museums, theatres, local cultural organisations, libraries, and many more work to keep culture alive online in difficult times. This category maps a selection of initiatives and organised events accessable online.
 represents various cultural opportunities available online in EU cities. Travel and social restrictions caused by the Covid-19 pandemic recently limited our possibility to physically visit popular European cultural venues. %and to taste cultural fragrances of European cities and towns in person. 
 As a reaction, several museums, theatres, local cultural organisations, libraries, and many more cultural venues organised various initiatives accessible online. This ontology class describes a set of performances and events organised to overcome physical distances. 

 \vskip0.3cm
 
 %\item 
 The \emph{Cinemas and theatres} class represents cultural venues such as cinemas, opera houses, theatres etc. 

\vskip0.3cm

 The \emph{Art galleries and museums} class models cultural venues such as art galleries and museums, while the \emph{Artworks} class describes 
 works of art produced and exhibited outside of the conventional gallery environment.

 \vskip0.3cm
 
 The \emph{Creative spaces} class depicts physical structures and elements at different scales that are deliberately designed to support creative work processes or to facilitate creativity.
 %%rephrased for ICICT
 %\item[$\rightarrow$] \emph{:Creative spaces} - Category that depicts physical objects and components at various scales that are intended to encourage or support creative business workflows and creativity.

 \vskip0.3cm
 
 The \emph{Historic sites} class represents historic sites, that is official locations where pieces of political, military, cultural, or social history have been preserved due to their cultural heritage value. Historic sites, also referred to as heritage sites, are usually protected by law, and many have been recognized with the official national historic site status. A historic site may be any building, landscape, site or structure that is of local, regional, or national significance (meaning approximately 50 years or older). 
 %%rephrased for ICICT
 %\item[$\rightarrow$] \emph{:Historic sites} - Set of official locations where historical artifacts from the realms of, among others, politics, war, culture, or society, have been preserved because of their cultural heritage value. Historic sites %, sometimes known as heritage sites, 
 %are frequently covered by legal protection, and many of them have received formal national historic site designation. Any building, location, landscape, or structure that is locally, regionally, or nationally significant, qualifies to be classified as a historic site. This often implies that the location must be at least 50 years old.

 \vskip0.3cm
 
 The \emph{Religious heritage} class defines spiritual and religious venues %, that is any form of property with religious or spiritual associations, 
 such as: churches, monasteries, crypts, shrines, sanctuaries, mosques, synagogues, temples, graveyards, sacred landscapes, sacred groves, and other landscape features, etc. 
 %%rephrased for ICICT
 %\item[$\rightarrow$] \emph{:Religious heritage} - Any type of property with religious or spiritual connotations, such as, among others, churches, sanctuaries, cemeteries, etc., falls under the category of religious heritage venues.

\vskip0.3cm

 The \emph{Memorials and monuments} class defines different types of historical attractions, while the \emph{Events and festivals} class describes the several kinds of events and celebrations.

\vskip0.3cm

 The \emph{Community spaces} class describes the various types of social and community areas.

\vskip0.3cm

 Finally, the \emph{Music venues} class models any location utilized for a concert or performance of music, including practice and recording studios.
  % venue used for a concert or musical performance, including recording and rehearsal studios.

\vskip0.3cm

%Note also that the 
Geometric and spatial data have been modelled using ArCo's location module. 
As a result, a gem can be composed of more places, which are described by the \emph{a-loc:LocationType} class in the ontology. 
In addition, a gem can be associated with a defined timespan. %Furthermore, it can be the case that a cultural location of a gem is valid only within a specific time interval. 
The time dimension is designed by adopting the \emph{a-loc:TimeIndexedTypeLocation} class in the ontology, expanding the \emph{TimeIndexedSituation} ODP. %ontology pattern. %\footnote{TimeIndexedSituation Pattern:\url{http://www.ontologydesignpatterns.org/cp/owl/timeindexedsituation.owl}} 

\vskip0.3cm

The ontology definition namespace is \href{https://jeodpp.jrc.ec.europa.eu/ftp/jrc-opendata/CC-COIN/cultural-gems/ontology-definition/cultural-gems.owl}{https://culturalgems.jrc.ec.europa.eu/ontology/cultural-gems/}, containing a total of 67 classes to date. 

\vskip0.3cm

\emph{Cultural gems} data populating the application are modelled as individuals assigned to a class of the designed ontology definition. The resulting data ontology accounts for around 2.9M triples to date. It is publically available in both Terse RDF Triple Language (Turtle) and RDF/XML formats \cite{JRCdataCG}.
The data ontology namespace \href{https://jeodpp.jrc.ec.europa.eu/ftp/jrc-opendata/CC-COIN/cultural-gems/ontology-individuals/cultural-gems-resources.owl}{https://culturalgems.jrc.ec.europa.eu/resource/}. %Content negotiation for the data is still not available but will be implemented soon. 

%Currently, 
Data resources are currently linked, whenever is appropriate, tp DBpedia \cite{DBpedia} and GeoNames \cite{GeoNames} with \textbf{owl:sameAs} relationships. The ontology is continuously expanding with additional alignments to other public cultural ontologies, %datasets %freely accessible online and 
%specialised to the cultural heritage domain, 
such as Europeana and ArCo data, which will come in the future. %, is envisaged. 

\vskip0.3cm

%Note that 
Ontology data names are expressed in lowercase, and eventual space characters have been replaced by dashes, as per common style guidelines for naming and labeling ontologies \cite{gangemi2009ontology}. Differently, ontology definition class names are expressed in uppercase, following common style guidelines used for naming and labeling ontology definitions. 

\vskip0.3cm

Both the definition and data ontologies are available %in RDF/XML and Turtle formats 
\cite{JRCdataCG} within the Joint Research Centre Data Catalogue \cite{JRCDataCatalogue} at the following permanent location:
\url{https://data.jrc.ec.europa.eu/dataset/9ee32efe-af81-48e4-8ad6-a0db06802e03}, and within the European Data portal \cite{EUDataPortal}, the official data repository for European data, at the following permanent location: \url{https://data.europa.eu/data/datasets/9ee32efe-af81-48e4-8ad6-a0db06802e03}.

\vskip 0.5cm

\textbf{Experimental design, materials and methods}\\
\vskip0.3cm 

Ontologies need to be appropriately created, maintained, adjusted, and expanded, as required. For these purposes, the developed ontology has been managed by adopting the \textit{Prot\'eg\'e} %\footnote{Prot\'eg\'e: \url{http://protege.stanford.edu}} 
ontology editor. \textit{Prot\'eg\'e} provides a fully integrated interface allowing to add, for example, new classes, or to update already existing ones. 
%To design our ontology, we rely on ontology design patterns (ODPs) principles \cite{gangemi2009ontology}, %Gangemi2005262
%that is, whenever possible we reuse existing ODPs from online ontology repositories. 

Ontology design patterns (ODPs) principles \cite{gangemi2009ontology} %Gangemi2005262
have been used in the modelled ontology. Whenever appropriate, existing ODPs from popular ontologies have been reused using OPLa ontology \cite{Asprino2021}. %Shimizu201823
For example, we have linked to classes and properties from the %available
OntoPia %Ontology
Public Administration vocabulary %\cite{OntoPia}
%
%the core %(roles, agents, locations) 
%modules of OntoPiA\footnote{OntoPia Ontology, \url{https://github.com/italia/daf-ontologie-vocabolari-controllati/tree/master/}, %and controlled vocabulary network for Italian Public Administration, and 
and from Cultural-ON, a popular ontology modelling cultural locations and events \cite{Lodi2017}. %\cite{CulturalON} 
%We also indirectly reuse patterns from existing ontologies, e.g. from the CIDOC Conceptual Reference Model (CRM) \cite{CIDOCCRM}, and include explicit alignments to them. % within ArCo.
The adoption of OPLa for annotating the reuse of other ontologies supports the overall class identification and mapping of the overall ontology.
%\vskip0.3cm

\vskip0.3cm

\emph{Cultural gems} data in the application populates the CG ontology
%are mapped as individuals of the CG ontology 
via an ETL (Extract-Transform-Load) nightly job developed in Python. Data are modelled as individuals of the ontology definition. 
These individuals are linked to data of other external ontologies by means of LIMES \cite{LIMES}, a popular link discovery tool for LOD. 

\vskip0.3cm

The ontology definition is stored in the RDF named-graph: 
\href{https://jeodpp.jrc.ec.europa.eu/ftp/jrc-opendata/CC-COIN/cultural-gems/ontology-definition/cultural-gems.owl}{https://culturalgems.jrc.ec.europa.eu/ontology/cultural-gems/},
while ontology resources %Data %resources
are stored in the RDF named-graph: \href{https://jeodpp.jrc.ec.europa.eu/ftp/jrc-opendata/CC-COIN/cultural-gems/ontology-individuals/cultural-gems-resources.owl}{https://culturalgems.jrc.ec.europa.eu/resource/}.

\vskip0.3cm

For programmers, %%rephrased for ICICT
both ontology definition and data are interrogable via SPARQL queries on a dedicated Virtuoso %\footnote{OpenLink Virtuoso: \url{https://virtuoso.openlinksw.com/}}
triplestore. For this purpose, we are leveraging CELLAR \cite{Cellar}, the official triplestore of the Publications Office of the European Union. %'s common repository of metadata and content, 
Queries are possible via the dedicated REST APIs available to the Virtuoso endpoint of CELLAR. % services. %\footnote{\url{https://data.europa.eu/data/datasets/sparql-cellar-of-the-publications-office}}. 
Queries can be made by editing the text area available in the interface for the SPARQL query language. SPARQL is the standard language reference and a W3C recommendation for querying RDF data. 
%
%%The SPARQL endpoint is also accessible as a REST web service; 
%The SPARQL endpoint\footnote{CELLAR Sparql endpoint: \url{https://publications.europa.eu/webapi/rdf/sparql}} is also accessible as a REST web service. It requires as input a user-defined SPARQL query and produces as output the query result in one of the following formats: \emph{text/html}, \emph{text/rdf +n3}, \emph{application/xml}, \emph{application/json}, or \emph{application/rdf+xml}. %%rephrased for ICICT
The REST APIs to the CELLAR triplestore \cite{Cellar} %\footnote{CELLAR Sparql endpoint: \url{https://publications.europa.eu/webapi/rdf/sparql}} 
need a specific query formatted in the SPARQL language as input, giving as output the query result in \emph{text/html},
\emph{text/rdf +n3}, \emph{application/xml},
\emph{application/json}, or \emph{application/rdf+xml}. %\footnote{Details on the synopsis will be please see: \url{http://wit.istc.cnr.it/prisma/webcontent/services.html}}.
The specific format is defined by the user as desired.

\vskip0.3cm

CONSTRUCT, ASK, DESCRIBE, SELECT queries can be performed to access the CG ontology. A CONSTRUCT query returns an RDF source constructed by substituting variables. An ASK query returns true or false indicating whether a query pattern matches or not. A SELECT query returns the variables bound in a query pattern match. %This type of query consists of three parts in general:
%1 PREFIX declares prefixes used in the query,
%2 SELECT identifies the variables to appear in the query results,
%3 WHERE provides the basic graph pattern to match against the date.

\vskip0.3cm

For example, we are given the \href{https://culturalgems.jrc.ec.europa.eu/map/4/139885}{Ancienne Belgique - AB} gem, a popular concert hall for contemporary music located in Brussels, Belgium, and the related resource in the data ontology corresponds to 
\href{https://publications.europa.eu/webapi/rdf/sparql?default-graph-uri=&query=select+distinct+*+from+%3Chttps%3A%2F%2Fculturalgems.jrc.ec.europa.eu%2Fresource%2F%3E+where+%7B%3Chttps%3A%2F%2Fculturalgems.jrc.ec.europa.eu%2Fresource%2Fcultural-gems%2F139885%3E+%3Fp+%3Fo%7D+&format=text%2Fhtml
}{https://culturalgems.jrc.ec.europa.eu/resource/cultural-gems/139885}.\\ 
To obtain all the RDF information related to this resource from the CELLAR endpoint, the corresponding SPARQL query is:\\
\vskip0.3cm
\noindent
\textit{SELECT * WHERE {$<$https://culturalgems.jrc.ec.europa.eu/resource/cultural-gems/139885$>$ ?p ?o}}\\
\vskip0.3cm

\noindent
In simple words, it translates into the selection of all the data triples having properties ?p and objects ?o, matching with the resource ``Ancienne Belgique - AB'' specified as subject.
%that, if executed in CELLAR, would produce the result available \href{https://publications.europa.eu/webapi/rdf/sparql?default-graph-uri=&query=DESCRIBE+%3Chttps%3A%2F%2Fculturalgems.jrc.ec.europa.eu%2Fresource%2Fcultural-gems%2F139885%3E&format=text%2Fhtml&timeout=0&debug=on&run=+Run+Query+}{here}.
\vskip0.3cm

The ontology can be also exploited by using the \emph{Live OWL Documentation Environment (LODE)} \cite{LODE}, a service that visualizes classes, object properties, data properties, named individuals, annotation properties, general axioms and namespace declarations from an RDF/OWL ontology, in a \href{http://150.146.207.114/lode/extract?url=https%3A%2F%2Fjeodpp.jrc.ec.europa.eu%2Fftp%2Fjrc-opendata%2FCC-COIN%2Fcultural-gems%2Fontology-definition%2Fcultural-gems.owl&lang=en}{human-readable HTML page}.
%human-readable HTML pages\footnote{\url{http://ec2-54-154-163-28.eu-west-1.compute.amazonaws.com:8090/lode/extract?url=https://jeodpp.jrc.ec.europa.eu/ftp/jrc-opendata/CC-COIN/cultural-gems/ontology-definition/cultural-gems.owl}}.
This tool enables user-oriented navigation and provides an intuitive description of elements of the ontology.
%
%It is also visualizable by \emph{WebVOWL}
%\footnote{WebVOWL 0.3, \url{http://vowl.visualdataweb.org/webvowl.html}} as a force-directed graph layout
%\footnote{\url{https://service.tib.eu/webvowl/\#iri=https://jeodpp.jrc.ec.europa.eu/ftp/jrc-opendata/CC-COIN/cultural-gems/ontology-definition/cultural-gems-skeleton.owl}}. 
%Both tools %, still available in the development framework only, 
%enable user-oriented visualizations and provide an intuitive description of elements of the ontology. Interaction techniques allow the user to better explore the ontology and customize the visualization. 
\vskip0.3cm

We also enable IRI dereferentiation by using 
%integrate a further visualization by means of %LodView and LodLive. %\footnote{LodView: \url{http://lodview.it/} and LodLive: \url{http://lodlive.it/}}
LodView \cite{LodView}, a Java web application compliant with W3C standards for content negotiation.
%which is able to offer a W3C standard compliant IRI dereferentiation. 
LodView improves the end user's experience by providing an HTML page describing the %based representations of our RDF resources. 
ontology data. Besides its intuitive interface, LodView adds interesting features to the content negotiation, such as the possibility to download the selected resource in different formats (\textit{xml, ntriples, turtle, ld+json}). For example, the \href{https://culturalgems.jrc.ec.europa.eu/map/4/139885}{Ancienne Belgique - AB} gem can be visualized via LodView at \href{https://lodview.it/lodview/?IRI=https%3A%2F%2Fculturalgems.jrc.ec.europa.eu%2Fresource%2Fcultural-gems%2F139885&sparql=https%3A%2F%2Fpublications.europa.eu%2Fwebapi%2Frdf%2Fsparql&prefix=https%3A%2F%2Fculturalgems.jrc.ec.europa.eu%2Fresource%2Fcultural-gems%2F139885}{Ancienne Belgique - AB lodview} (Figure \ref{lodview-ancienne}). With this representation, users can explore and navigate the resource, noting e.g. that this gem is a subclass of the \href{https://lodview.it/lodview/?IRI=https%3A%2F%2Fculturalgems.jrc.ec.europa.eu%2Fontology%2Fcultural-gems%2FMusicVenue&sparql=https%3A%2F%2Fpublications.europa.eu%2Fwebapi%2Frdf%2Fsparql&prefix=https%3A%2F%2Fculturalgems.jrc.ec.europa.eu%2Fontology%2Fcultural-gems%2FMusicVenue}{Music Venue} class, and has associated a main \href{www.abconcerts.be}{source} homepage. It is also possible to download the raw information of the resource in various formats, e.g. \href{https://publications.europa.eu/webapi/rdf/sparql?output=application%2Fodata%2Bjson&query=DESCRIBE+%3Chttps://culturalgems.jrc.ec.europa.eu/resource/cultural-gems/139885%3E}{json} or \href{https://publications.europa.eu/webapi/rdf/sparql?output=text%2Fcsv&query=DESCRIBE+%3Chttps://culturalgems.jrc.ec.europa.eu/resource/cultural-gems/139885%3E}{csv}, among others.

\begin{figure}[!ht]
\begin{center}
 \fboxsep=0.5mm%padding thickness
 \fboxrule=0.5pt%border thickness
 \fcolorbox{black}{white}{\includegraphics[width=0.7\linewidth]{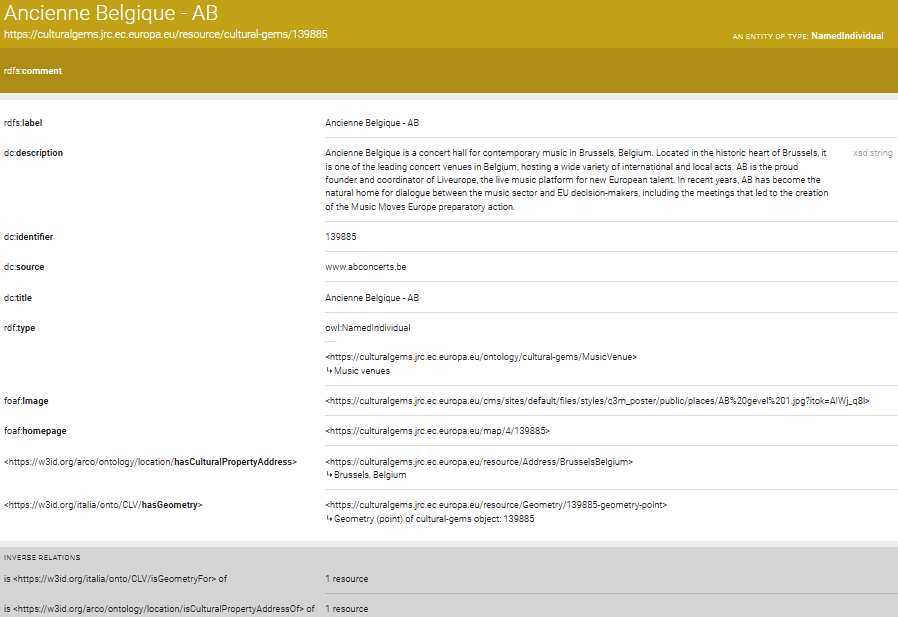}}
	\caption{IRI dereferentiation of the \href{https://culturalgems.jrc.ec.europa.eu/map/4/139885}{Ancienne Belgique - AB} gem by means of LodView.}
	\label{lodview-ancienne}
\end{center}
\end{figure}

\vskip0.3cm

We further integrate LodLive \cite{LodLive} to facilitate the navigation of the data in the ontology. LodLive enables the visualization of the resources of an ontology as an efficient graph. %LodLive is a visualization tool which allows to depict the RDF data via an effective graph representation. 
The user can, for example, examine the structure of the RDF/OWL data and interactively explore the relationships of a particular ontology individual. 
%The user is able, for instance, to expand the relationships of a given ontology resource in an automated way and explore the structure of the RDF data. 
LodLive enables also the association of geo-coordinates and images to the data resources and the computation of inverse and \textbf{owl:sameAs} links.
%It is also possible to associate images and geo-coordinates to the data instances, and evaluate \textbf{owl:sameAs} and inverse relations. 
As an example, 
Figure \ref{lodlive-ancienne} shows how the \href{https://culturalgems.jrc.ec.europa.eu/map/4/139885}{Ancienne Belgique - AB} gem is interconnected with the other ontology individuals by means of the LodLive graph visualization.

\begin{figure}[!ht]
\begin{center}
 \fboxsep=0.5mm%padding thickness
 \fboxrule=0.5pt%border thickness
 \fcolorbox{black}{white}{\includegraphics[width=0.70\linewidth]{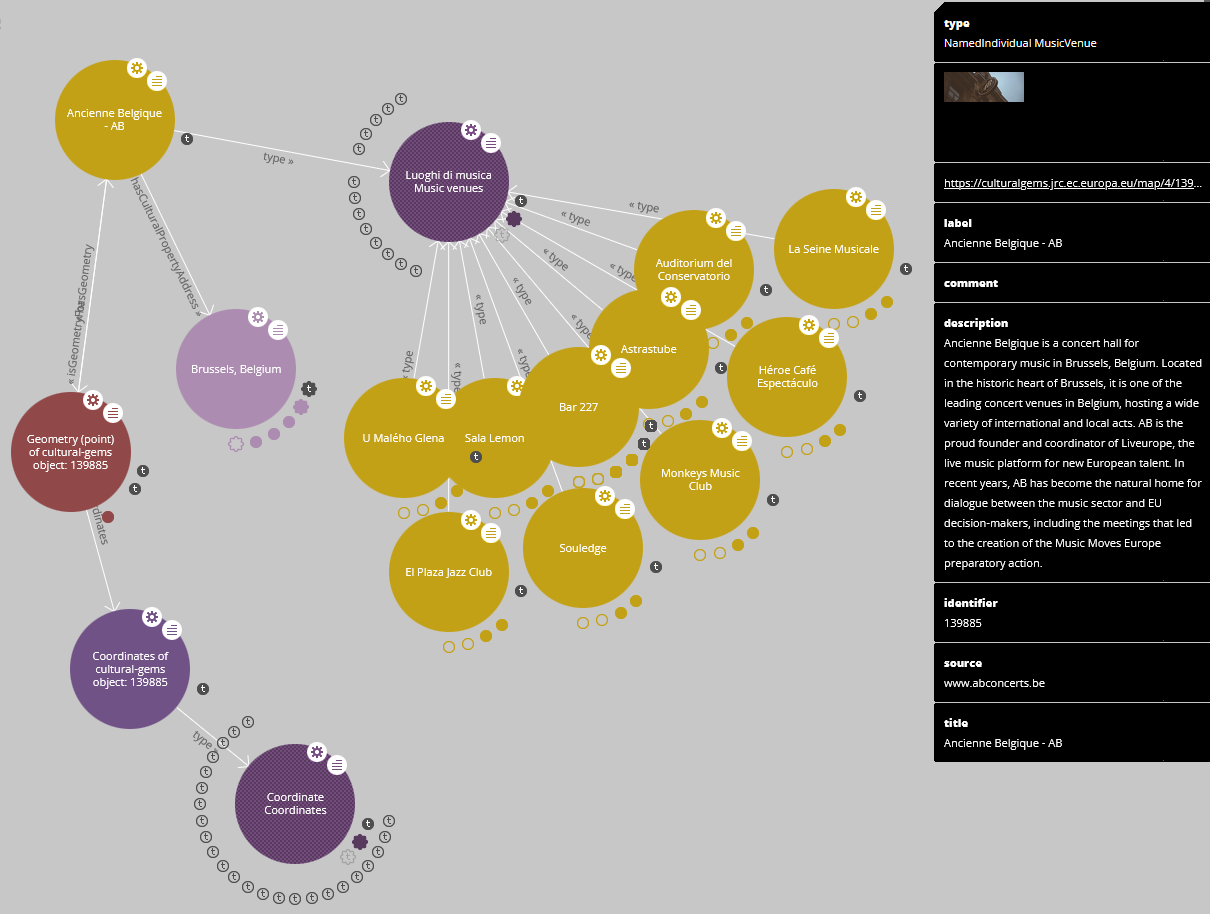}}
	\caption{Graph visualization of the \href{https://culturalgems.jrc.ec.europa.eu/map/4/139885}{Ancienne Belgique - AB} gem by means of LodLive. 
 % View of the relationships of the \href{https://culturalgems.jrc.ec.europa.eu/map/4/139885}{Ancienne Belgique - AB} gem to the other data entities using LodLive.
 }
	\label{lodlive-ancienne}
\end{center}
\end{figure}

\vskip 0.5cm

\textbf{Ethics statements}\\
\vskip0.3cm
The present work did not involve human subjects, animals or information from social media platforms.
\vskip0.5cm

\noindent
\textbf{CRediT author statement}\\
\vskip0.3cm
\noindent
\textbf{Sergio Consoli}: Conceptualization, Data curation, Methodology, Software, Supervision, Writing- Original draft preparation. \textbf{Valentina Alberti}: Investigation, Visualization, Supervision, Writing- Reviewing and Editing. \textbf{Cinzia Cocco}: Software, Investigation, Validation, Writing- Reviewing and Editing. \textbf{Francesco Panella}: Data curation, Investigation, Supervision, Writing- Reviewing and Editing. \textbf{Valentina Montalto}: Investigation.\\
\vskip0.5cm

\textbf{Acknowledgments}\\
\vskip0.3cm
We would like to thank the colleagues of the Competence Centre on Composite Indicators and Scoreboards (COIN) at the Joint Research Centre of the European Commission for their support. The views expressed are purely those of the authors and may not in any circumstance be regarded as stating an official position of the European Commission. \\
\vskip0.3cm
The authors would like to thank also the CELLAR team at the Publications Office of the European Union for their support on the use of their public knowledge base for hosting our ontology and interacting with it.\\
\vskip0.3cm
This research did not receive any specific grant from funding agencies in the public, commercial, or nonprofit sectors.\\
\vskip0.5cm

\textbf{Declaration of interests}\\
\vskip0.3cm
\begin{itemize}
\item[X]{The authors declare that they have no known competing financial interests or personal relationships that could have appeared to influence the work reported in this paper.}

\item[$\square$]{The authors declare the following financial interests/personal relationships which may be considered as potential competing interests: }
\end{itemize}
\vskip0.3cm

\bibliographystyle{elsarticle-num}
\bibliography{bibliofile.bib}

\end{flushleft}
\end{document}